\documentclass{article}

\usepackage{amssymb,amsfonts,amsmath,stmaryrd}
\usepackage{cite,enumerate,float,indentfirst}
\usepackage{color}
\usepackage{tikz}
\usetikzlibrary{arrows,snakes,backgrounds}
\usepackage{url}
\usepackage{hyperref}
\usepackage{textcomp}
\usepackage{amsthm}

\def\be{\begin{eqnarray}}
\def\ee{\end{eqnarray}}
\def\nn{\nonumber}

\def\tr{{\rm tr}\,}
\def\Tr{{\rm Tr}\,}

\def\Schur{{\rm Schur}}
\def\Schur{S}
\def\sgn{{\rm sgn}}

\definecolor{red}{rgb}{1,0,0}
\definecolor{orange}{rgb}{1,0.5,0}
\definecolor{violet}{rgb}{0.7,0,1}

\hoffset= - 1.0in         % switch off for draft style
\begin{document}

\begin{center}
\begin{small}
\hfill MIPT/TH-19/21\\
\hfill FIAN/TD-18/21\\
\hfill ITEP/TH-33/21\\
\hfill IITP/TH-23/21\\

\end{small}
\end{center}

\vspace{.5cm}

\begin{center}
\begin{Large}\fontfamily{cmss}
\fontsize{15pt}{27pt}
\selectfont
	\textbf{Natanzon-Orlov model and refined superintegrability}
	\end{Large}
	
\bigskip \bigskip

\begin{large}
A. Mironov$^{a,b,c,}$\footnote{mironov@lpi.ru; mironov@itep.ru},
V. Mishnyakov$^{d,a,c,e,}$\footnote{mishnyakovvv@gmial.com},
A. Morozov$^{d,b,c,}$\footnote{morozov@itep.ru},
A. Zhabin$^{d,c,}$\footnote{alexander.zhabin@yandex.ru}
\end{large}

\bigskip

\begin{small}
$^a$ {\it Lebedev Physics Institute, Moscow 119991, Russia}\\
$^b$ {\it Institute for Information Transmission Problems, Moscow 127994, Russia}\\
$^c$ {\it ITEP, Moscow 117218, Russia}\\
$^d$ {\it MIPT, Dolgoprudny, 141701, Russia}\\
$^e$ {\it Institute for Theoretical and Mathematical Physics, Lomonosov Moscow State University, 119991 Moscow, Russia}
\end{small}
 \end{center}

\bigskip

\begin{abstract}
We reconsider
the simple matrix model description of Hurwitz numbers
proposed by S. Natanzon and A. Orlov, which uses
the superintegrability property of
the complex matrix model, and discuss a way of
its possible supersymmetric extension to approach spin Hurwitz numbers.
\end{abstract}

\bigskip

\section{Introduction}

In \cite{MM}, it was suggested to reduce
remarkable properties \cite{UFN3} of matrix models \cite{Mehta}
to their superintegrability,
that is, to the fact that averages of symmetric functions/characters
are again expressed through the same symmetric functions/characters (see many examples in
\cite{DiF}--\cite{Zabz}, and also some preliminary results in \cite{Kaz}--\cite{MKR})
depending only on the type of matrix model and on the matrix size.

In the simplest possible case of complex matrix model \cite{MMcomp},
where the integration goes over $N_1\times N_2$ matrices $Z$,
the claim was \cite{MM}
\be\label{si}
\left<S_R\{\Tr (ZZ^\dagger)^k\}\right>:=
\int S_R\{\Tr (ZZ^\dagger)^k\} \cdot\exp\left(-\Tr ZZ^\dagger\right) d^2Z \ =\
{  S_R\{N_1\}\ S_R\{N_2\}\over d_R}
\ee
Here $S_R\{p_k\}$ is the Schur polynomial (which is defined to be a symmetric function $S_R(z_i)$ of the  variables $z_i$, or a graded polynomial $S_R\{p_k\}$
of the power sums $p_k:=\sum_iz_i^k$).
At the l.h.s., the role of $z_i$ is played by the eigenvalues of the matrix $ZZ^\dagger$, we will use the notation $S[ZZ^\dagger]$ in such cases. At the r.h.s., $p_k=N_1$ or $N_2$; $d_R:=S_R\{\delta_{k,1}\}$.
The integration measures is normalized so that $<1>\,:=1$.

It is natural to look for a further generalization (refinement)
of this formula, where $N_1$, $N_2$ are substituted by arbitrary square matrices,
and, indeed, such formula can be found \cite{Ivan1,Ivan2,NO,MMkon}:
\begin{equation} \label{mba}
\boxed{
    \Big< S_{R}[AZBZ^{\dagger}] \Big> = \frac{S_{R}[A] S_{R}[B]}{d_{R}}
    }
\end{equation}
and leads to a remarkable matrix model
representation of Hurwitz numbers by
S.Natazon and A.Orlov \cite{NO}.
Here $A$ and $B$ are arbitrary square matrices of the sizes $N_1\times N_1$
and $N_2\times N_2$ respectively.
In our opinion, this provides the complete form of the
superintegrability relation for  complex matrix model,
and one should seek for its generalization to other matrix models,
including Hermitian model and its fermionic counterpart,
governing the spin Hurwitz numbers.

In this short note, we review the application of (\ref{mba}) to Hurwitz numbers in \cite{NO},
and provide the combinatorial proof of this relation both in the bosonic and fermionic cases along the lines of \cite{MMkon}.

 \section{Application to Hurwitz numbers (NO model)}

Generating function of the Hurwitz numbers is made our of the Schur polynomials:
\be
{\cal Z}_m\{p^{(1)},\ldots,p^{(m)}\}
=\sum_{R} d_R^{2-m}\cdot \Schur_R\{p^{(1)}\}\ldots \Schur_R\{p^{(m)}\}
\label{genH}
\ee
In \cite{NO}, a matrix model realization of (\ref{genH}) was proposed,
\be
{\cal Z}_m\{p^{(1)},\ldots,p^{(m)}\}
= \prod_{i=1}^m \int d^2Z_i e^{-\tr Z_iZ_i^\dagger} \exp\left( \tr \prod_{i=1}^m Z_iA_i\right)
\exp\left( \tr \prod_{i=1}^m Z_i^\dagger\right)
= \nn \\
= \prod_{i=1}^m \int  d^2Z_i e^{-\tr Z_iZ_i^\dagger}
\exp\Big( \tr A_1Z_2A_2Z_3A_3 \ldots Z_3^\dagger Z_2^\dagger\Big)
\ee
with $p^{(i)}_k = \Tr A_i^k$.
Evaluating of the integral at the r.h.s. of this expression is
done using at the first step the elementary property
\be\label{5}
\int  e^{\tr A_1Z_2U_2Z_2^\dagger} e^{-\tr Z_2 Z_2^\dagger } d^2Z_2
= \frac{1}{{\rm Det}\big(I\otimes I - A_1\otimes U_2\big)}=
\exp\left( \sum_{k} \frac{ \tr A^{k} \tr U_2^{k}}{k} \right)
= \sum_R \Schur_R\{\overbrace{p^{(1)}_k}^{\tr A_1^k}\}\cdot
\Schur_R[U_2]
\ee
where $U_2=A_2Z_3A_3 \ldots Z_3^\dagger $, and further iterating with the superintegrability property (\ref{mba}): at the second step, integrating (\ref{5}) over $Z_3$ makes out of $\Schur_R[U_2]$ the product $\displaystyle{{\Schur_R\{p^{(2)}_k\}\over d_R}}\cdot
\Schur_R[U_3]$ with $U_3=A_3Z_4A_4 \ldots Z_4^\dagger $, etc. The final result is exactly (\ref{genH}).

\section{Towards $Q$-Schur functions}

An important goal is to find a generalization of the NO models to reproduce the spin Hurwitz numbers. The corresponding generating function is given, instead of (\ref{genH}) by \cite{MMN}
\be
\mathfrak{Z}^p_m\{p^{(1)},\ldots,p^{(m)}\}
=\sum_{R\in SP} (-1)^{p\cdot l_R}\cdot \mathfrak{d}_R^{2-m}\cdot Q_R\{p^{(1)}\}\ldots Q_R\{p^{(m)}\}
\label{gensH}
\ee
where the generating function additionally depends on the parameter $p=0,1$ that encodes the spin structure, the sum goes over the strict partitions with $l_R$ parts (i.e. over those with all parts distinct), $\mathfrak{d}_R=\displaystyle{1\over 2}Q_R\{\delta_{k,1}\}$, and $Q_R$ are the Schur $Q$-functions \cite{QS,MMN}. In other words, to deal with the spin Hurwitz numbers, one needs to find an extension of (\ref{mba}) from $\Schur_R$ to the
$Q$-Schur functions $Q_R$. To this end, one needs to add fermions already at the very first step.
In order to see this, note that the counterpart of \eqref{mba} is (see sections 4-5 below)
\begin{equation}\label{mfa}
    \boxed{
    \left< S_{R}[A \Psi B \Psi^{\dagger}] \right> = (-1)^{|R|} \frac{S_{R}[-A] S_{R}[B]}{d_{R}}
    }
\end{equation}
It nicely reproduces the known result
\begin{equation}
    \int d^{2} \Psi e^{-\tr \Psi \Psi^{\dagger} + \tr A\Psi B\Psi^{\dagger}} = \sum_{R} (-1)^{|R|} S_{R}[-A] S_{R}[B] = \exp\left( -\sum_{k} \frac{(-1)^{k} \tr A^{k} \tr B^{k}}{k} \right) = \det(1+ A\otimes B)
\end{equation}
Now one easily gets for the superintegration:
\be
\int  e^{\tr AZBZ^\dagger+\tr A\Psi B\Psi^\dagger}
e^{-\tr Z Z^\dagger+\tr \Psi\Psi^\dagger } d^2Z d^2\Psi=
\frac{{\rm Det}\big(I\otimes I + A\otimes B\big)}
{{\rm Det}\big(I\otimes I - A\otimes B\big)}=\exp\left( 2\sum_{k\ odd} \frac{ \tr A^{k} \tr B^{k}}{k} \right)=\nn\\
= \sum_R Q_R\{\tr A^k\}Q_R\{\tr B^k\}
\ee
Thus, the Schur $Q$-functions emerge when one adds fermions.

However, a $Q$-counterpart of (\ref{mba}) should not be found in this way,
because the theory of $Q$-functions seems to generalize the Hermitian
rather than the complex matrix model, i.e. the superintegrability relation is
\be
\left<S_R\{\Tr X^k\}\right>:=
\int S_R\{\Tr X^k\} \cdot\exp\left(-{1\over 2}\Tr X^2\right) dX =
{S_R\{N\}\ S_R\{\delta_{k,2}\}\over d_R}
\ee
instead of (\ref{si}).
It agrees with the fact that the $Q$-functions are
characters \cite{MMZh} of representations of the supergroup associated with a special reduction of $gl(n|n)$, of the queer algebra
$\mathfrak{q}(n)$ \cite{qn}, while
the characters associated with $gl(n|m)$ are the supersymmetric Schur functions \cite{sS}, which, in terms of the power sums $p_k$'s are reduced to the usual
Schur functions.
We address this story in a separate publication.

\section{A proof}

In this section we provide a formal proof of (\ref{mba}) and (\ref{mfa}).
It is based on the Wick theorem \cite[Eq.(7)]{MMten}
\begin{equation}\label{Wick_bosonic}
    \left< \prod_{i=1}^{n} Z_{a_{i} \alpha_{i}} \overline{Z}_{b_{i} \beta_{i}} \right> = \sum_{\gamma \in S_{n}} \prod_{i=1}^{n} \delta_{a_{i}}^{b_{\gamma(i)}} \delta_{\alpha_{i}}^{\beta_{\gamma(i)}}
\end{equation}
let $|R| = |\Delta| = n$, then
\begin{equation}\label{schur_via_p}
    \left< S_{R}[AZBZ^{\dagger}] \right> = \sum_{\Delta \vdash n} \frac{\psi_{R}(\Delta)}{z_{\Delta}} \cdot \langle p_{\Delta} \rangle
\end{equation}
where
\begin{equation}
    \langle p_{\Delta} \rangle = \left< \prod_{i=1}^{l(\Delta)} \tr(AZBZ^{\dagger})^{\Delta_{i}} \right> = \prod_{i=1}^{n} A_{a_{i} k_{i}} B_{\alpha_{i} m_{i}} \cdot \left< \prod_{i=1}^{n} Z_{k_{i} \alpha_{i}} \overline{Z}_{a_{\sigma(i)} m_{i}} \right>
\end{equation}
here $\sigma$ is any element from the conjugacy class $\Delta$. Then
\begin{equation}
    \langle p_{\Delta} \rangle = \prod_{i=1}^{n} A_{a_{i} k_{i}} B_{\alpha_{i} m_{i}} \sum_{\gamma \in S_{n}} \prod_{i=1}^{n} \delta_{k_{i}}^{a_{\gamma(\sigma(i))}} \delta_{\alpha_{i}}^{m_{\gamma(i)}} = \sum_{\gamma \in S_{n}} \prod_{i=1}^{n} A_{a_{i} a_{\gamma(\sigma(i))}} B_{m_{\gamma(i)} m_{i}} = \sum_{\gamma \in S_{n}} A_{\gamma \circ \sigma} B_{\gamma}
\end{equation}
where the last equation means that if $\gamma \circ \sigma$ and $\gamma$ are in the conjugacy classes $\lambda$ and $\mu$ respectively, then
\begin{equation}
    A_{\gamma \circ \sigma} = A_{\lambda} = \prod_{i=1}^{l(\lambda)} \tr A^{\lambda_{i}}, \;\;\;\;\;\; B_{\gamma} = B_{\mu} = \prod_{i=1}^{l(\mu)} \tr B^{\mu_{i}}
\end{equation}
and \eqref{schur_via_p} has the form
\begin{equation}\label{schur_via_matrices}
    \left< S_{R}[AZBZ^{\dagger}] \right> = \sum_{\Delta \vdash n} \frac{\psi_{R}(\Delta)}{z_{\Delta}} \cdot \sum_{\gamma \in S_{n}} A_{\gamma \circ \sigma} B_{\gamma}
\end{equation}
Now we use the formula
\begin{equation}
    p_{\Delta} = \sum_{R \vdash n} \psi_{R}(\Delta) S_{R}(p)
\end{equation}
to rewrite \eqref{schur_via_matrices}:
\begin{equation}
    \begin{gathered}
    \langle S_{R}[AZBZ^{\dagger}] \rangle = \sum_{\Delta \vdash n} \frac{\psi_{R}(\Delta)}{z_{\Delta}} \cdot \sum_{\gamma \in S_{n}} \left( \sum_{R_{1} \vdash n} \psi_{R_{1}}(\gamma \circ \sigma) S_{R_{1}}[A] \right) \left( \sum_{R_{2} \vdash n} \psi_{R_{2}}(\gamma) S_{R_{2}}[B] \right) = \\
    = \sum_{\Delta, R_{1}, R_{2}} \frac{\psi_{R}(\Delta)}{z_{\Delta}} S_{R_{1}}[A] S_{R_{2}}[B] \left( \sum_{\gamma \in S_{n}} \psi_{R_{1}}(\gamma \circ \sigma) \psi_{R_{2}}(\gamma) \right)
    \end{gathered}
\end{equation}
The generalized orthogonality relations \cite[Eq.(9)]{MMten}
\begin{equation}
    \sum_{\gamma \in S_{n}} \psi_{R_{1}}(\gamma \circ \sigma) \psi_{R_{2}}(\gamma) = \frac{\psi_{R_{1}}(\sigma)}{d_{R_{1}}} \delta_{R_{1}, R_{2}}
\end{equation}
and $\psi_{R}(\sigma) = \psi_{R}(\Delta)$ imply that
\begin{equation}
    \langle S_{R}[AZBZ^{\dagger}] \rangle = \sum_{R_{1}} \frac{S_{R_{1}}[A] S_{R_{1}}[B]}{d_{R_{1}}}  \underbrace{\left( \sum_{\Delta} \frac{\psi_{R}(\Delta) \psi_{R_{1}}(\Delta)}{z_{\Delta}} \right)}_{\delta_{R, R_{1}}} = \frac{S_{R}[A] S_{R}[B]}{d_{R}}
\end{equation}
This completes the proof of (\ref{mba}).

%\end{proof}

\bigskip

A simple, still useful corollary is
\begin{equation}\label{p_id}
    \langle p_{[1^{n}]} \rangle = \langle (\tr AZBZ^{\dagger})^{n} \rangle = \sum_{\gamma \in S_{n}} A_{\gamma \circ id} B_{\gamma} = \sum_{\Delta \vdash n} [\Delta] A_{\Delta} B_{\Delta}
\end{equation}
where $[\Delta] = \frac{n!}{z_{\Delta}}$ is the number of elements in the conjugacy class $\Delta$.
\begin{equation}
    \begin{gathered}
    \sum_{R \vdash n} S_{R}[A] S_{R}[B] = \sum_{R \vdash n} \sum_{\Delta_{1}} \frac{\psi_{R}(\Delta_{1})}{z_{\Delta_{1}}} A_{\Delta_{1}} \sum_{\Delta_{2}} \frac{\psi_{R}(\Delta_{2})}{z_{\Delta_{2}}} B_{\Delta_{2}} = \sum_{\Delta_{1} \Delta_{2}}  \frac{A_{\Delta_{1}} B_{\Delta_{2}}}{z_{\Delta_{1}}} \underbrace{\left( \sum_{R \vdash n} \frac{\psi_{R}(\Delta_{1}) \psi_{R}(\Delta_{2})}{z_{\Delta_{2}}} \right)}_{\delta_{\Delta_{1}, \Delta_{2}}} =\\
    = \sum_{\Delta} \frac{1}{z_{\Delta}} A_{\Delta} B_{\Delta} = \frac{1}{n!} \sum_{\Delta} [\Delta] A_{\Delta} B_{\Delta}
    \end{gathered}
\end{equation}
Thus we can rewrite \eqref{p_id} as
\begin{equation}
    \langle p_{[1^{n}]} \rangle = \langle (\tr AZBZ^{\dagger})^{n} \rangle = \sum_{\Delta \vdash n} [\Delta] A_{\Delta} B_{\Delta} = n! \sum_{R} S_{R}[A] S_{R}[B]
\end{equation}

\section{Implications  for fermionic averages}

Now in order to get (\ref{mfa}), a fermionic counterpart of (\ref{mba}), one needs the fermionic Wick theorem:
\begin{equation}\label{Wick_fermionic}
    \boxed{
    \left< \prod_{i=1}^{n} \Psi_{a_{i} \alpha_{i}} \overline{\Psi}_{b_{i} \beta_{i}} \right> = \sum_{\gamma \in S_{n}} \sgn(\gamma) \prod_{i=1}^{n} \delta_{a_{i}}^{b_{\gamma(i)}} \delta_{\alpha_{i}}^{\beta_{\gamma(i)}}
    }
\end{equation}
where $\sgn(\gamma) = \pm 1$ depends on the parity of permutation:
\begin{equation}
    \begin{array}{cc}
        \gamma & \sgn(\gamma) \\
        \left[1\right] & 1 \\
        \left[2\right] & -1 \\
        \left[1,1\right] & 1 \\
        \left[3\right] & 1 \\
        \left[2,1\right] & -1 \\
        \left[1,1,1\right] & 1 \\
        \left[4\right] & -1 \\
        \left[3,1\right] & 1 \\
        \left[2,2\right] & 1 \\
        \left[2,1,1\right] & -1 \\
        \left[1,1,1,1\right] & 1
    \end{array}
\end{equation}
An explicit formula for $\sgn(\gamma)$ is
\begin{equation}
    \sgn(\gamma) = (-1)^{\sum_{i} (\gamma_{i} - 1)} = (-1)^{|\gamma| - l(\gamma)} = (-1)^{|\gamma| + l(\gamma)}
\end{equation}
Now literally repeating the derivation of (\ref{mba}), we arrive at (\ref{mfa}).

Again, the simplest check is
\begin{equation}
    \begin{gathered}
    \langle p_{[1^{n}]} \rangle = \left< (\tr A\Psi B \Psi^{\dagger})^{n} \right> = \sum_{\gamma \in S_{n}} \sgn(\gamma) A_{\gamma \circ id} B_{\gamma} = \sum_{\Delta \vdash n} (-1)^{|\Delta| + l(\Delta)} [\Delta] A_{\Delta} B_{\Delta} = (-1)^{n} \sum_{\Delta} (-A)_{\Delta} B_{\Delta} = \\
    = n! (-1)^{n} \sum_{R \vdash n} S_{R}[-A] S_{R}[B]
    \end{gathered}
\end{equation}

\section{Conclusion}

In this note, we revised the extension (\ref{mba})
of the superintegrability property
for the simplest and archetypical complex matrix model
 \cite{Ivan1,Ivan2,NO,MMkon}, which gives rise to the matrix model description
 of the Hurwitz numbers  by S. Natanzon and A. Orlov \cite{NO}.
We came across this description by reexamining the legacy
of Sergey Natanzon. This remarkable result adds to the variety of his contributions
to modern mathematical physics. We discussed a possible extension
to fermionic models in order to approach to the spin Hurwitz numbers.

\section*{Acknowledgements}

The work is partly  supported by the Russian Science Foundation
(Grant No.20-12-00195).


\begin{thebibliography}{12}

\bibitem{MM} A.~Mironov, A.~Morozov,
  %``On the complete perturbative solution of one-matrix models,''
  Phys.\ Lett.\ {\bf B771} (2017) 503,
%  doi:10.1016/j.physletb.2017.05.094
arXiv:1705.00976

\bibitem{UFN3} A. Morozov,
Phys.Usp.(UFN) {\bf 37} (1994) 1;
hep-th/9502091; hep-th/0502010\\
A. Mironov, Int.J.Mod.Phys. {\bf A9} (1994) 4355; Phys.Part.Nucl.
{\bf 33} (2002) 537; hep-th/9409190

\bibitem{Mehta} J. Ginibre,
%Statistical ensembles of complex, quaternion and real matrices.”,
J. Math. Phys. {\bf 6} (1965) 440\\
M.L. Mehta, {\it Random Matrices,} 2.ed.,
Academic Press, 1990

\bibitem{DiF} P.~Di Francesco, C.~Itzykson, J.~B.~Zuber,
  %``Polynomial averages in the Kontsevich model,''
  Commun.\ Math.\ Phys.\  {\bf 151} (1993) 193,
  %doi:10.1007/BF02096753
  hep-th/9206090
  
\bibitem{IdF} P.~Di Francesco, C.~Itzykson,
%``A Generating function for fatgraphs,''
Ann. Inst. H. Poincare Phys. Theor. \textbf{59} (1993) 117-140,
hep-th/9212108

\bibitem{Ivan1} I.~K.~Kostov, M.~Staudacher,
%``Two-dimensional chiral matrix models and string theories,''
Phys. Lett. \textbf{B394} (1997) 75-81,
%doi:10.1016/S0370-2693(96)01664-4
hep-th/9611011

\bibitem{Ivan2} I.~K.~Kostov, M.~Staudacher, T.~Wynter,
%``Complex matrix models and statistics of branched coverings of 2-D surfaces,''
Commun. Math. Phys. \textbf{191} (1998) 283-298,
%doi:10.1007/s002200050269
hep-th/9703189

\bibitem{AMMN} A. Alexandrov, A. Mironov, A. Morozov, S. Natanzon,
JHEP 11 (2014) 080,  arXiv:1405.1395

\bibitem{Orlov}  S. Natanzon, A. Orlov, arXiv:1407.8323

\bibitem{PSh} C.~Cordova, B.~Heidenreich, A.~Popolitov, S.~Shakirov,
  %``Orbifolds and Exact Solutions of Strongly-Coupled Matrix Models,''
  Commun.\ Math.\ Phys.\  {\bf 361} (2018)   1235,
 % doi:10.1007/s00220-017-3072-x
  arXiv:1611.03142

  \bibitem{IMM} H.~Itoyama, A.~Mironov, A.~Morozov,
  %``Ward identities and combinatorics of rainbow tensor models,''
  JHEP {\bf 1706} (2017) 115,
% doi:10.1007/JHEP06(2017)115
arXiv:1704.08648

  \bibitem{MMten} A.~Mironov, A.~Morozov,
  %``Correlators in tensor models from character calculus,''
  Phys.\ Lett.\ {\bf B774} (2017) 210,
 % doi:10.1016/j.physletb.2017.09.063
arXiv:1706.03667

\bibitem{MPS} A.~Morozov, A.~Popolitov and S.~Shakirov,
  %``On (q,t)-deformation of Gaussian matrix model,''
  Phys.\ Lett. {\bf B784} (2018) 342,
  %doi:10.1016/j.physletb.2018.08.006
  arXiv:1803.11401

    \bibitem{MMsum} A.~Mironov, A.~Morozov,
  %``Sum rules for characters from character-preservation property of matrix models,''
  JHEP {\bf 1808} (2018) 163,
%doi:10.1007/JHEP08(2018)163
arXiv:1807.02409

\bibitem{MMell}   A.~Mironov, A.~Morozov,
%``Elliptic $q,t$ matrix models,''
Phys. Lett. \textbf{B816} (2021), 136196,
%doi:10.1016/j.physletb.2021.136196
arXiv:2011.01762; ibid.,
%``Towards elliptic deformation of $q,t$-matrix models,''
136221,
%doi:10.1016/j.physletb.2021.136221
arXiv:2011.02855

\bibitem{MMkon} A.~Mironov, A.~Morozov,
%``Superintegrability of Kontsevich matrix model,''
Eur. Phys. J. \textbf{C81} (2021) 270,
%doi:10.1140/epjc/s10052-021-09030-x
arXiv:2011.12917

\bibitem{MMl} A.Mironov, A.Morozov,  Phys.Lett. {\bf B816} (2021) 136268,  arXiv:2102.01473

\bibitem{Zabz} L.~Cassia, R.~Lodin, M.~Zabzine,
%``On matrix models and their $q$-deformations,''
JHEP \textbf{10} (2020) 126,
%doi:10.1007/JHEP10(2020)126
arXiv:2007.10354

\bibitem{Kaz} V.A. Kazakov, M. Staudacher, T. Wynter, %Advances in Large N Group Theory
%and the Solution of Two-Dimensional $R^2$ Gravity, 
hep-th/9601153, 1995 Carg\`ese Proceedings

\bibitem{Ramg} S. Corley, A. Jevicki, S. Ramgoolam, Adv.Theor.Math.Phys. {\bf 5} (2002) 809-839, hep-th/0111222

\bibitem{KPSS} C. Kristjansen, J. Plefka, G. W. Semenoff, M. Staudacher, Nucl.Phys. {\bf B643} (2002) 3-30, hep-th/0205033

\bibitem{BEM} M. Tierz, Mod. Phys. Lett. A19 (2004) 1365-1378, hep-th/0212128\\
A. Brini, B. Eynard, M. Mari\~no, Annales Henri Poincar\'e. Vol. 13. No. 8. SP Birkh\"{a}user Verlag Basel, 2012, arXiv:1105.2012

\bibitem{MKR} R. de Mello Koch, S. Ramgoolam, arXiv:1002.1634

\bibitem{MMcomp} T. Morris, %{\it Checkered surfaces and complex matrices},
Nucl.Phys. {\bf b356} (1991) 703-728\\
Yu. Makeenko, Pis'ma v ZhETF {\bf 52} (1990) 885-888\\
Yu. Makeenko, A. Marshakov, A. Mironov, A. Morozov, Nucl.Phys. {\bf B356} (1991) 574-628

\bibitem{NO}
S.Natanzon, A.Orlov,
%``Hurwitz numbers from Feynman diagrams,''
Teor. Mat. Fiz. \textbf{204} (2020) 396-429
[erratum: Theor. Math. Phys. \textbf{205} (2020) 1546],
%doi:10.1134/S0040577920090068
arXiv:2006.07396

\bibitem{MMN} A.~Mironov, A.~Morozov, S.~Natanzon,
  %``Cut-and-join structure and integrability for spin Hurwitz numbers,''
  Eur.\ Phys.\ J.\ {\bf C80} (2020)  97,
  %doi:10.1140/epjc/s10052-020-7650-2
arXiv:1904.11458

\bibitem{QS} I. Schur,
%{\it Uber die Darstellung der symmetrischen und der alternierenden Gruppe durch
%gebrochene lineare Substitutionen},
J. Reine Angew. Math. {\bf 139} (1911) 155-250\\
I.G. Macdonald,
{\it Symmetric functions and Hall polynomials}, Second Edition, Oxford University Press,
1995

\bibitem{MMZh} A.~Mironov, A.~Morozov, A.~Zhabin,
%``Spin Hurwitz theory and Miwa transform for the Schur Q-functions,''
arXiv:2111.05776

\bibitem{qn} Cheng, Shun-Jen, Weiqiang Wang, {\em ``Dualities and Representations of Lie Superalgebras."}, 2012

\bibitem{sS}  A. Berele, A. Regev,
%Hook Young diagrams with applications to combinatorics and to representations of Lie superalgebras,
Advances in Mathematics {\bf 64} (1987) 118-175\\
P. Pragacz, A. Thorup,
%On a Jacobi–Trudi identity for supersymmetric polynomials,
Adv. Math. {\bf 95} (1992) 8–17\\
E.M. Moens, J. van der Jeugt,
%A determinantal formula for supersymmetric Schur polynomials,
J. Algebraic Combin. {\bf 17} (2003) 283–30

\end{thebibliography}
\end{document}